\documentclass[aps,prl]{revtex4}

\usepackage{graphicx}
\usepackage[dvips]{color}

\def\be{\begin{equation}}
\def\ee{\end{equation}}
\def\ba{\begin{array}}
\def\ea{\end{array}}
\def\bea{\begin{eqnarray}}
\def\eea{\end{eqnarray}}

\begin{document}


\title{Crack propagation through phase separated glasses: \\effect of
the characteristic size of disorder}

\author{Davy Dalmas$^1$, Anne Lelarge$^1$ and Damien Vandembroucq$^{1,2}$}
\affiliation
{$^1$ Unit\'e Mixte CNRS/Saint-Gobain ``Surface du Verre et Interfaces''\\
39 Quai Lucien Lefranc, 93303 Aubervilliers cedex, FRANCE\\
$^2$ Laboratoire PMMH, UMR 7636 CNRS/ESPCI/P6/P7\\
10 rue Vauquelin 75231 Paris cedex 05, FRANCE
}

\begin{abstract}
We perform fracture experiments on nanoscale phase separated glasses
and measure crack surface roughness by atomic force microscopy. The
ability of tuning the phase domain size by thermal treatment allows us
to test thoroughly the predictions of crack font depinning models
about the scaling properties of crack surface roughness. It appears
that in the range of validity of these depinning models developed for
the fracture of brittle materials, our experimental results show a
quantitative agreement with theoretical predictions: beyond the
characteristic size of disorder, the roughness of crack surfaces obeys
the logarithmic scaling early predicted by Ramanathan, Ertas  and
Fisher\cite{Ramanathan-PRL97b}.
\end{abstract}

\maketitle



 Ever since an early paper by Mandelbrot {\it et al}
\cite{Mandelbrot84Nat} about the fractal character of crack surfaces,
a growing interest has developed in the understanding of fracture of
heterogeneous
materials\cite{Herrmann-Roux-book90,EBouchaud-JPC97,Alava-AdP06}.  In
particular many
studies\cite{Mandelbrot84Nat,EBouchaud90EPL,Maloy-PRL92,EBouchaud-JPC97,Schmittbuhl-PRL97,Boffa-EPJAP98,Hinojosa-JMR02,Bonamy-PRL06,Ponson-PRL06b}
have dealt with the statistical characterization of both crack front
geometry and crack surface roughness. It appears that, over a broad
spectrum of length scales, these objects obey a self-affine symmetry:
they are left statistically invariant rescaling with different ratios
depending on the direction. More specifically, self-affinity implies
that the typical height differences $\Delta h$ along the surface
scales as $\Delta h \propto \Delta x^\zeta$ when measured over a
distance $\Delta x$, $\zeta$ being a real parameter called the
roughness, or Hurst, exponent. Early studies proposed that the
roughness of various fractured materials, independently of their
ductile\cite{EBouchaud90EPL} or brittle\cite{Maloy-PRL92} nature,
could be described over a wide range of scales with a unique value of
the roughness exponent $\zeta \simeq 0.8$.

\begin{figure}
\begin{center}
\includegraphics [height=0.22\textwidth]{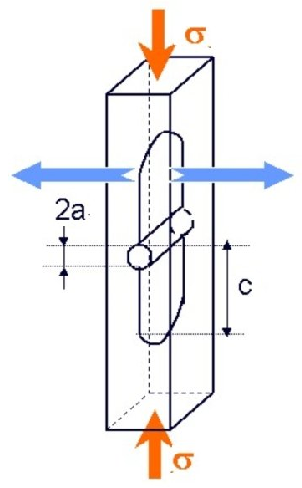} \hspace{0.5cm}
\includegraphics [height=0.25\textwidth]{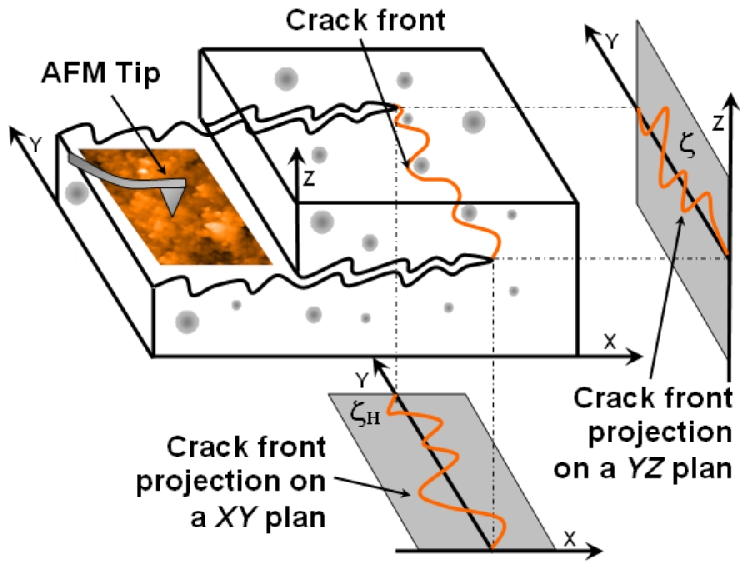} \hspace{1.0cm}
\includegraphics [height=0.22\textwidth]{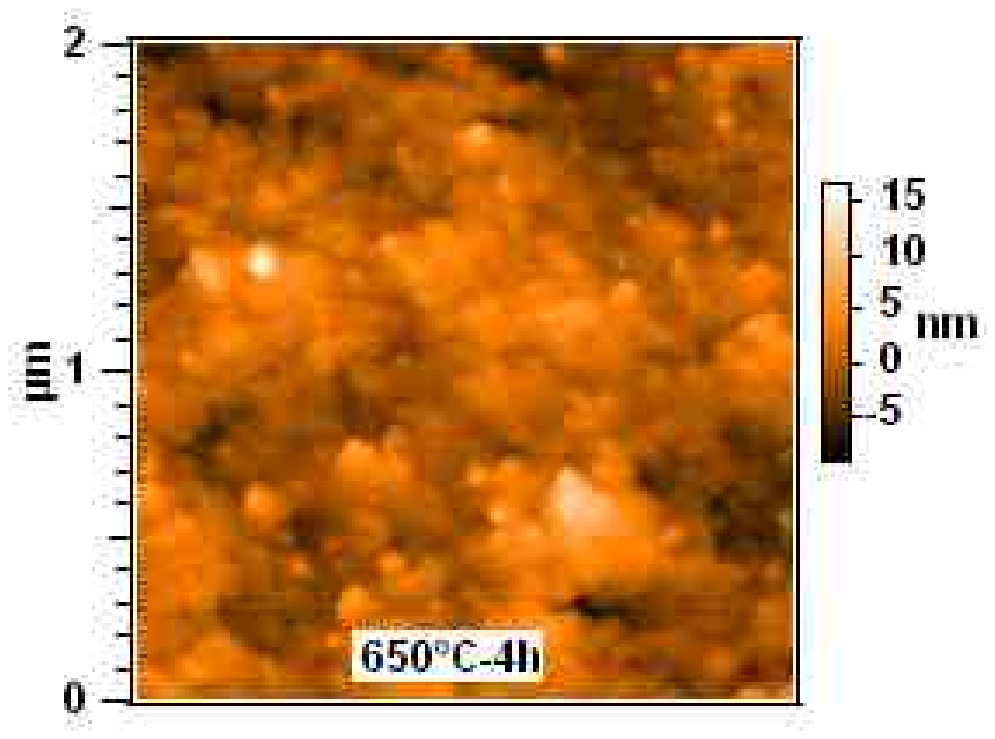}
\caption{{\bf Experimental methods.} Left:
geometry of the DCDC mechanical test. Two longitudinal cracks
propagate from the cylindrical hole. Center: sketch of a propagating
crack front pinned by heterogeneities (here figured as grey dots). The
front develops a roughness both in the out-of-plane direction, visible
on the $y-z$ plane projection and in the direction of propagation,
visible on the $x-y$ plane projection. When scaling invariant, this
roughness is characterized by an exponent $\zeta$ in the out-of-plane
direction and an exponent $\zeta_H$ in the direction of
propagation. AFM measurements are performed on the crack
surface. Right: AFM image of a surface obtained after fracture of a
glass sample annealed at $650^\circ$C over 4h.}\label{set-up}
\end{center}
\end{figure}

This apparent universality motivated the development of
models\cite{Schmittbuhl-PRL95,Ramanathan-PRL97b} based on the
depinning of an elastic line through a random
landscape\cite{Kardar-PR98} and the description of
long range elastic interactions along a rough crack
front\cite{GaoRice-JAM89,Willis-JMPS97}. A crack surface can indeed be
considered as the trace left by a crack front propagating through a
disordered material\cite{Daguier-PRL97} (see Fig. \ref{set-up}a for a
representation of crack front propagation).  In such a description, a
particular site pinned by a microscopic tough zone may overcome the
obstacle with the help of the additional elastic force due to the
straining of the structure around the obstacle.  This competition
between a roughening effect due to the quenched random environment and
a smoothing effect due to elastic interactions along the front leads
to a rich phenomenology. 
When the Stress Intensity Factor $K$ remains
below a certain threshold, $K^*$ the front can only
propagate over a finite length and then stops; above threshold,
the front can move at a finite, but highly fluctuating,
velocity. Moreover, this transition appears to be critical, with a
collection of critical exponents (among which the roughness exponent
of the elastic front) ruling its behavior close to threshold, with the
strong property of universality coming naturally with this
observation.

These depinning models thus intend to characterize the
associated critical  transition and to capture the geometry and the
intermittent dynamics of the crack front.
The case of interfacial plane cracks was first
developed\cite{Schmittbuhl-PRL95} to study the in-plane roughness
({\it i.e.} in the direction of propagation) of the front in the
propagation plane. The more complex case of a 3D crack front was then
discussed\cite{Ramanathan-PRL97b}, allowing the study of the
out-of-plane roughness ({\it i.e.} in the direction normal to the mean
plane of propagation) responsible for the fracture surface roughness.


Though qualitatively correct, the results appeared to be
quantitatively rather disappointing since these models predict only a
logarithmic scaling of the crack surface's out-of-plane roughness
\cite{Ramanathan-PRL97b}, {\it i.e.}  the roughness exponent predicted
by the depinning model was $\zeta^{dep}=0$ (here and in the
following the notation $\zeta^{dep}$ will refer to the exponents
predicted by the depinning models while the notation $\zeta$ will
refer to experimental estimates) whereas  the experimental estimate was
$\zeta\simeq 0.8$. Even in the simpler case of plane crack
propagation, the roughness exponent characterising the in-plane
roughness of the crack front was found to be significantly lower in
models ($\zeta_H^{dep}\simeq 0.4$) \cite{Rosso-PRE02,VR-PRE04} than in
experiments ($\zeta_H \simeq 0.55$) \cite{Schmittbuhl-PRL97}.


We argue here within that the main reason for the apparent discrepancy
between experimental results and scaling predictions of depinning
models stems from the dubious status in most experiments of a key
hypothesis of the models: a clear separation between the scale of
disorder and the scale of measurement.  To test this argument, we
present results of fracture experiments performed on a series of model
heterogeneous brittle materials: phase separated
glasses\cite{DLV-JNCS07}.  Thermal treatments of various durations
allowed the fine tuning of phase domain sizes. Phase separated glasses
thus  differ only by one single parameter, the lengthscale of their
microstructure, with all other material properties remaining
identical. Such an experimental procedure thus allows us to study the
dependence of the spatial extent of the scaling regime on the
characteristic size of the internal disorder.

\begin{figure}
\begin{center}
\includegraphics [height=0.24\textwidth]{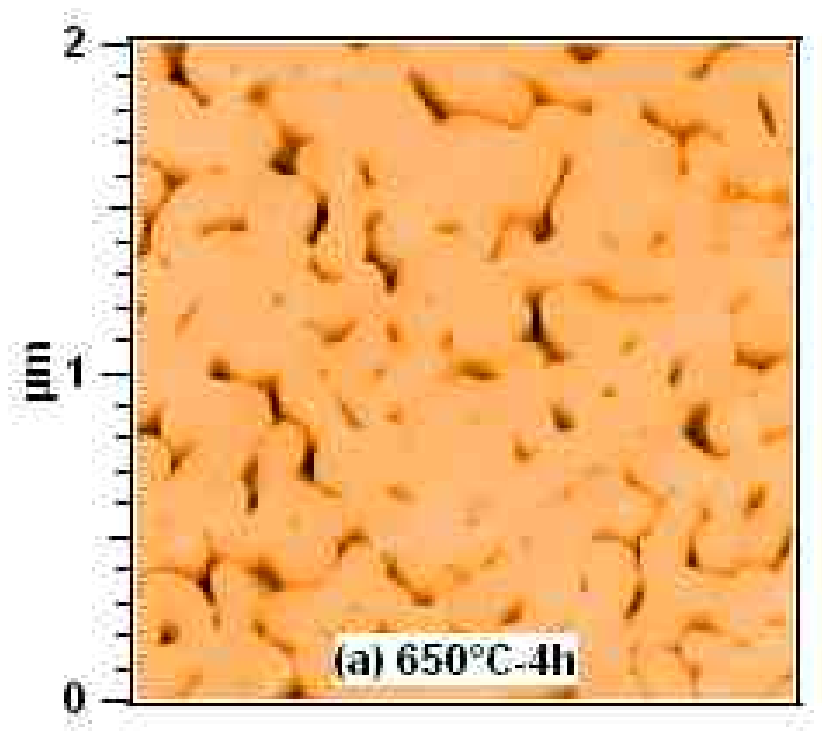} \hspace{0.5cm}
\includegraphics [height=0.24\textwidth]{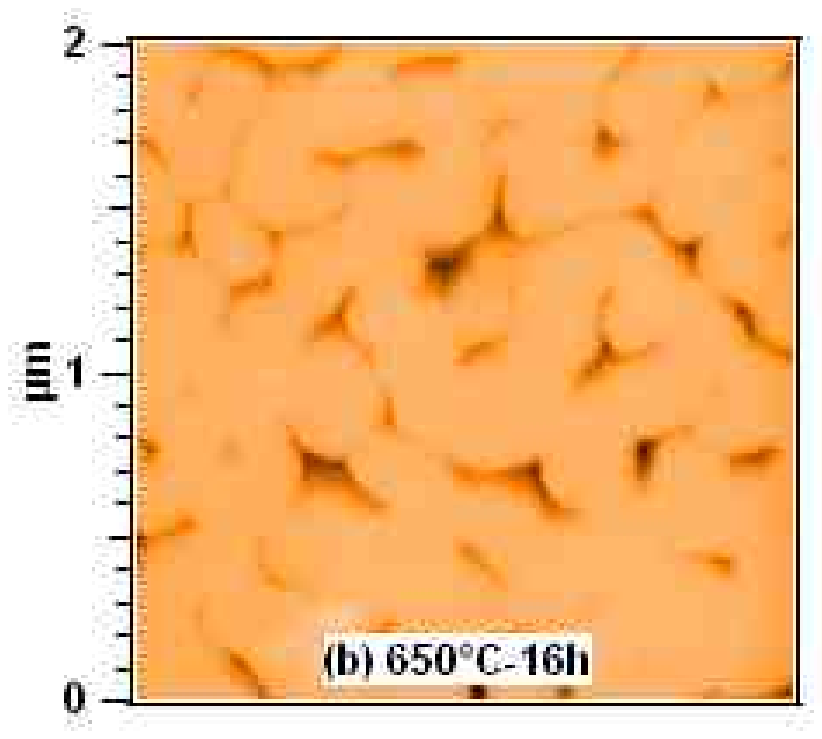}\hspace{0.5cm}
\includegraphics [height=0.24\textwidth]{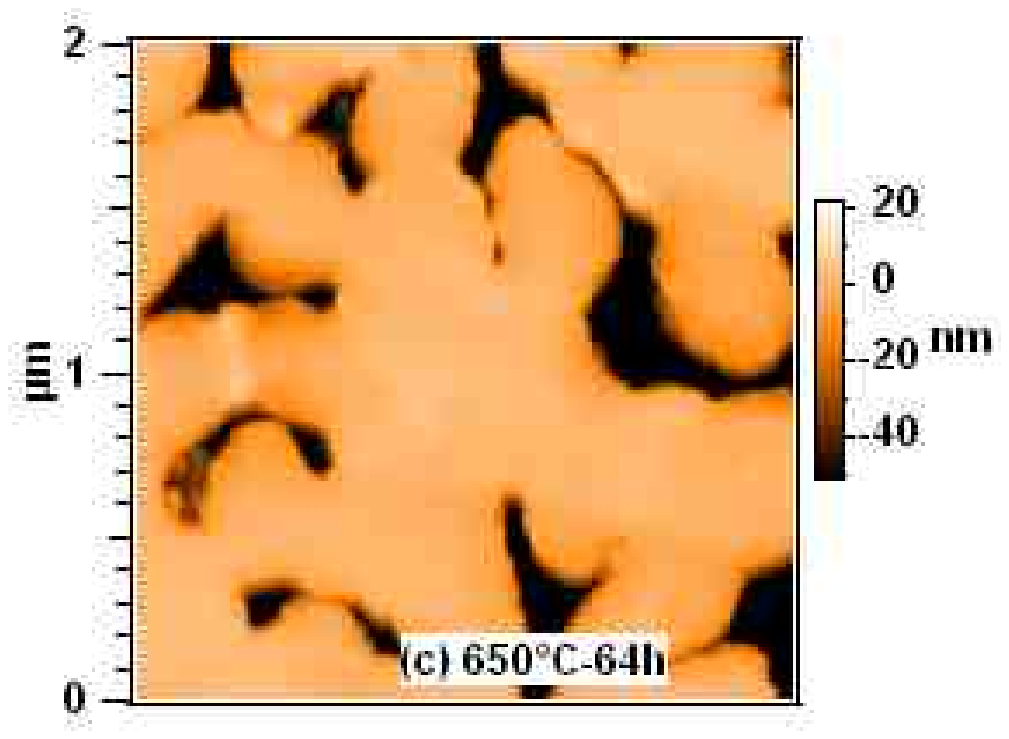}
\end{center}
\caption{{\bf AFM images obtained after acid etching on glass samples
annealed at $650^\circ$C over 4h, 16h and 64h respectively}. Under
annealing, alkali borosilicate separate into a near pure silica
phase and another phase concentrating all other components. The
latter phase can be eliminated by acid etching, the height contrast
obtained on AFM images thus reveals the phase contrast (see also
Ref. \cite{DLV-JNCS07}).}
\label{montafm}
\end{figure}

Phase separated glasses have been prepared from an alkali borosilicate
glass of composition (in weight): SiO$_2$ 70\%, B$_2$O$_3$ 25\%,
Na$_2$O 2.5\% and K$_2$O 2.5\%.  The raw materials were mixed together
and melted at 1550$^\circ$C and the melt was then refined over 2 hours
in order to obtain a homogeneous glass; after quenching, the glass was
annealed for 1h at 630$^\circ$C in order to relax the internal thermal
stresses. At this stage, phase separation had already initiated,
attested to by the slightly opalescent character of the glass. The
major phase is near-pure silica while the minor phase concentrates
other constituents. Heat treatments at $650^\circ$C over increasing
durations (4h, 16h, 64h) were then applied to increase more and more
the size of phase domains as shown in Fig. \ref{montafm} and
summarised in Table \ref{domain_size}. It was thus possible to prepare
glasses with controlled phase domain sizes, ranging from around 20 to
100 nm.  Further details concerning the preparation and the characterisation of the glass
samples  can be found in
Ref. \cite{DLV-JNCS07}. 
\begin{table}
\centering
\begin{tabular}{|c|c|c|c|}
\hline
Heat treatment duration  &4h &16h &64h \\
\hline
Phase domain size $d$ &$28.2\pm2.5$ nm &$56.3\pm9.3$ nm &$92.2\pm9.4$nm\\
\hline
\end{tabular}
\caption{Size $d$ of phase domains (estimated by the correlation
length at half height) {\it vs} duration of thermal treatment
at 650$^\circ$C. See Ref. \cite{DLV-JNCS07} for quantitative details
about the kinetics of phase separation on these
glasses.}\label{domain_size}
\end{table}


Stable dynamic fracture experiments on these heterogeneous materials
were performed by using a DCDC set-up (Double Cleavage Drilled
Compression) (see Fig. \ref{set-up}-Left) and a statistical analysis
of the fracture surface roughness carried out from Atomic Force
Microscopy (AFM) images (see Fig. \ref{set-up}-Center for the
principle and Fig. \ref{set-up}-Right for an AFM image of roughness
masurements obtained on a fracture surface).  From the above
mentioned glass, parallepipedic shaped samples
(5mm$\times$5mm$\times$25mm) with a central hole 1mm in diameter have
then been prepared. Stable cracks were then propagated in these by
using DCDC mechanical test.  Roughness measurements, coming from
height images (TM-AFM) were then immediately performed in areas lying
between two crack arrest lines.  The measurements are thus
performed in the immediate vicinity of the propagation threshold: $K
\approx K^*$.

The surface morphology of all samples
has been characterised by AFM height measurement in tapping mode
(TM-AFM), using a Nanoscope III A from Digital Instruments with Al
coated tip (BudgetSensors - model BS-Tap 300 Al). The Al coating
thickness used was 30nm, the resonant frequency 300 kHz and the
stiffness constant 40N/m. Images were recorded at a scan frequency
between 0.8 and 1Hz for a resolution of $512\times512$ pixels.  For
each sample (i.e. for each thermal treatment conditions), at least 3
AFM images were performed for 4 different scan areas ranging between
$500\times 500$ nm$^2$ and $8\times 8$ $\mu$m$^2$.

\begin{figure}
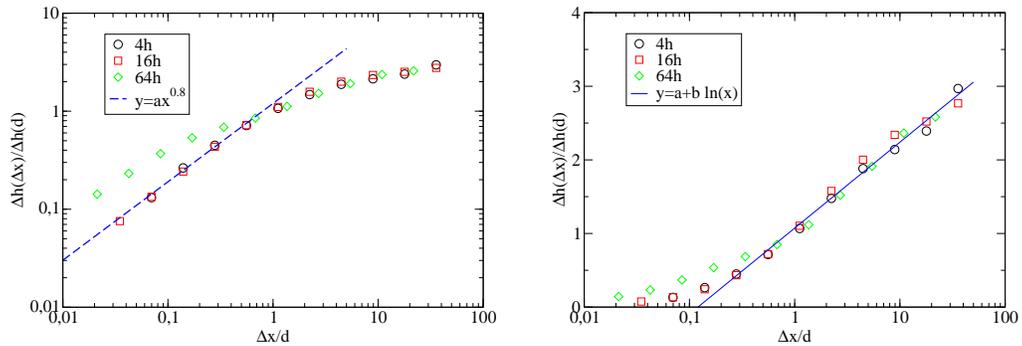

\begin{center}
   \includegraphics[height=0.25\textwidth]{stable-loglog.eps}
\hspace{0.5cm}
   \includegraphics[height=0.25\textwidth]{stable-linlog.eps}

  \caption{{\bf Roughness of fracture surfaces after rescaling by the phase
  domain size $d$ along the $x-$axis and the typical roughness $\Delta
  h(d)$ at size $d$ along the $y-$axis}. Left: in log-log scale an
  apparent self-affine regime with exponent $\zeta=0.8$ may be
  identified up to the the size of the phase domains. An indicative
  power law of exponent 0.8 is displayed. Right: in linear-log scale, a
  simple logarithmic regime appears to describe the surface roughness
  as predicted by depinning models. The straight line corresponds to
  a logarithmic fit performed on the 3 sets of data for lengths larger
  than the domain size.}\label{roughness}
\end{center}
\end{figure}

Beyond its imaging ability, AFM has recently been used as a truly
quantitative tool to study glass surfaces. It is, for example,
possible to obtain a quantitative validation of the description of
fused glass surfaces by frozen capillary waves\cite{SLSV-EPJB06} or of
the kinetics of phase separation in glasses\cite{DLV-JNCS07}. In both
cases, not only are the expected scaling regimes recovered but the
prefactors of the scaling laws can also be extracted and shown to be
consistent with the associated physical parameters (diffusivity,
interface tension).
In the present case of fracture surfaces, it appears that one can
identify an apparent scaling regime up to the size of the
heterogeneities (Fig. \ref{roughness}a). This result is consistent with the standard
analysis even if the scaling range  is limited.  More interestingly,
beyond the size of heterogeneities, one observes
(Fig. \ref{roughness}b) a clear logarithmic scaling as predicted by
the depinning model early proposed by Ramanathan {\it et al}
\cite{Ramanathan-PRL97b,Ramanathan-PhD98}.

More specifically Fig. \ref{roughness} represents the evolution of
height differences $\Delta h(\Delta x)=\left(\langle [h(x+\Delta x)-
h(x)]^2\rangle_x\right)^{1/2}$ measured on a scale $\Delta x$ on the
crack surface of each of the three types of sample. The results could
be collapsed onto a master curve after rescaling by the phase domain
size $d$ along the $x$-axis and by the typical height difference
$\Delta h(d)$ along the $y-$axis. When data are displayed on a log-log
scale (Fig. \ref{roughness}a), an apparent power law behaviour of
exponent $\zeta=0.8$ is identified, extending over one decade up to
the size of heterogeneity. Such a scaling law {\it below} the size of
heterogeneities, if genuine, could not be explained in the framework
of depinning models which predict scaling regimes only {\it above} the
typical scale of disorder. However if data are now represented on a
simple linear-log scale, we obtain above the characteristic scale of
the heterogeneities excellent agreement between our results and the
logarithmic scaling predicted by depinning models. Such a data
collapse after rescaling by the characteristic size of disorder
$\xi_0$ is fully consistent with a description of fracture propagation
as a critical transition. In such models the scaling regime is indeed
expected to hold in the spatial range $[\xi_0,\xi]$ where $\zeta_0$ is
the characteristic size of the toughness disorder $\xi \propto \xi_0
|K^*-K|^{-\nu}$ is the correlation length, diverging as a power law of
exponent $\nu$ when the stress intensity factor $K$ approaches the
critical threshold $K^*$ (which is nothing here but the effective
toughness of the material) and is directly proportional to $\xi_0$.
Here, propagation is performed close to threshold and
the scaling regime is thus expected to be wide enough to be easily
observed. Note that in case of a strongly overdriven crack, $K \gg
K^*$, dynamic instabilities set in and a roughness of entirely
different nature is expected to appear.  Depinning models thus seem to
give an excellent account of the scaling of the roughness of crack
surfaces obtained from the quasistatic fracture of brittle
heterogeneous materials such as the phase separated glasses used in
the present study.



The present results thus may contribute to clarify the current debate
about fracture and universality.  In view of the persistent
discrepancy between experimental measurements and model predictions,
efforts have indeed been made to either improve these depinning
models\cite{AKV-PRE06,Katzav-EPL06} or look for alternative scenarios
for fracture propagation\cite{Bonamy-PRL06,Bonamy-IJF06}. Higher order
terms were taken into account in the depinning model\cite{AKV-PRE06}
but apparently did not change the scaling
behaviour\cite{Bouchbinder-PRE07}. High velocity propagation was
discussed\cite{Katzav-EPL06} as a possible mechanism to anneal the
frozen toughness noise in the equation describing the front
propagation.  The interplay between local damage and crack propagation
was proposed but led to controversy at both the
theoretical\cite{Hansen-PRL03,Schmittbuhl-PRL03,Alava-PRL04,Schmittbuhl-PRL04}
and the experimental\cite{Celarie-PRL03,Guin-IJF06} levels.

Moreover, in light of the variety of experimental results, the initial
claim of universality itself may need to be tempered.  Beyond the
variability of sacling exponents estimated experimentally, when
measurements are performed on materials with a well defined
micro-structure such as glass\cite{Bonamy-PRL06}, sintered glass
beads\cite{Ponson-PRL06b} or some cast aluminium
alloys\cite{Hinojosa-JMR02}, it appears that either the scaling range
is very limited or that it does not extend beyond the grain size
whereas a truly universal scaling regime should extend beyond the
characteristic size of heterogeneities.

Within this context, the results of the present study suggest a simple
scenario. When considering the stable propagation of {\it brittle}
heterogeneous samples, depinning models can account quantitatively for
the scaling of crack surface roughness {\it beyond} the scale of
heterogenities. This means in particular that we need to clearly
separate the spatial scales within which the material can be
considered as brittle (beyond the size of the ``process zone'' where
dissipative mechanism associated with crack propagation take place)
from the range of scales within which damage or ductility take
place. Depinning models will not give any predictions below the size
of the process zone. It will thus be necessary to resort to
alternative models (stress-weighted
percolation\cite{Schmittbuhl-PRL03}, random fuse
models\cite{Alava-AdP06}) to account for fracture propagation in {\it
ductile} heterogeneous materials.

Note again that in the present framework of fracture, the notion
of disorder is rather subtle. The fluctuating quantity to consider is
actually the toughness of the material. However the toughness disorder
is not a direct reflection of the structural disorder but need be
``convoluted'' at the scale of the process zone. Two
lengthscales have thus to be considered, the size of the process zone
and the characteristic size of the structural disorder. In the present
case, glasses being very brittle, the typical size of the proces zone
does not overcome a few nanometers, below the size of the phase
domains and the characateristic scale of the toughness disorder can be
safely identified with the size of the structural disorder.

The common experimental observation of a wide scaling regime on
fracture surfaces with roughness exponents differing from the
depinning predictions may therefore result from the non-respect of the
hypothesis of depinning models: scale separation between toughness
disorder and experiments or perfect brittleness. This may be the case
for ductile materials for which the size of the process zone lies in
the spatial range of roughness measurements or very heterogeneous
materials presenting heterogeneities over a wide range of scales. In
particular the ``classical'' scaling regime characterized by a
roughness exponent $\zeta\simeq0.8$ may correspond to the existence of
additional microscopic mechanisms ({\it e.g.}  damage or other
non-linear behaviour) over a more limited range of scales. The present
results also suggest that experiments displaying a scaling regime only
up to the characteristic size of disorder may be revisited to test
whether they exhibit the logaritmic scaling of roughness expected
beyond that characteristic size.

This first quantitative validation of the depinning scenario to
describe brittle crack propagation through heterogeneous materials
thus provides a framework to extend the use of critical transitions
concepts to fracture, for instance in order to better predict
fluctuations and homegeneous behaviour from nano- to
macro-scale\cite{RVH-EJMA03}.

\medskip

\noindent {\bf Acknowledgements}

We acknowledge fruitful discussions with Daniel Bonamy, Elisabeth Bouchaud and St\'ephane Roux.

\bibliographystyle{apsrev}
\bibliography{depinning,vdb,BS,bibart2}   

\begin{thebibliography}{33}
\expandafter\ifx\csname natexlab\endcsname\relax\def\natexlab#1{#1}\fi
\expandafter\ifx\csname bibnamefont\endcsname\relax
  \def\bibnamefont#1{#1}\fi
\expandafter\ifx\csname bibfnamefont\endcsname\relax
  \def\bibfnamefont#1{#1}\fi
\expandafter\ifx\csname citenamefont\endcsname\relax
  \def\citenamefont#1{#1}\fi
\expandafter\ifx\csname url\endcsname\relax
  \def\url#1{\texttt{#1}}\fi
\expandafter\ifx\csname urlprefix\endcsname\relax\def\urlprefix{URL }\fi
\providecommand{\bibinfo}[2]{#2}
\providecommand{\eprint}[2][]{\url{#2}}

\bibitem[{\citenamefont{Ramanathan et~al.}(1997)\citenamefont{Ramanathan,
  Erta\c{s}, and Fisher}}]{Ramanathan-PRL97b}
\bibinfo{author}{\bibfnamefont{S.}~\bibnamefont{Ramanathan}},
  \bibinfo{author}{\bibfnamefont{D.}~\bibnamefont{Erta\c{s}}},
  \bibnamefont{and} \bibinfo{author}{\bibfnamefont{D.~S.}
  \bibnamefont{Fisher}}, \bibinfo{journal}{Phys. Rev. Lett.}
  \textbf{\bibinfo{volume}{79}}, \bibinfo{pages}{873} (\bibinfo{year}{1997}).

\bibitem[{\citenamefont{Mandelbrot et~al.}(1984)\citenamefont{Mandelbrot,
  Passoja, and Paullay}}]{Mandelbrot84Nat}
\bibinfo{author}{\bibfnamefont{B.~B.} \bibnamefont{Mandelbrot}},
  \bibinfo{author}{\bibfnamefont{D.~E.} \bibnamefont{Passoja}},
  \bibnamefont{and} \bibinfo{author}{\bibfnamefont{A.~J.}
  \bibnamefont{Paullay}}, \bibinfo{journal}{Nature}
  \textbf{\bibinfo{volume}{308}}, \bibinfo{pages}{721} (\bibinfo{year}{1984}).

\bibitem[{\citenamefont{Herrmann and Roux}(1990)}]{Herrmann-Roux-book90}
\bibinfo{author}{\bibfnamefont{H.}~\bibnamefont{Herrmann}} \bibnamefont{and}
  \bibinfo{author}{\bibfnamefont{S.}~\bibnamefont{Roux}},
  \emph{\bibinfo{title}{Statistical Models for the Fracture of Disordered
  Media}} (\bibinfo{publisher}{North-Holland}, \bibinfo{year}{1990}).

\bibitem[{\citenamefont{Bouchaud}(1997)}]{EBouchaud-JPC97}
\bibinfo{author}{\bibfnamefont{E.}~\bibnamefont{Bouchaud}},
  \bibinfo{journal}{J. Phys. Cond. Mat.} \textbf{\bibinfo{volume}{9}},
  \bibinfo{pages}{4319} (\bibinfo{year}{1997}).

\bibitem[{\citenamefont{Alava et~al.}(2006)\citenamefont{Alava, Nukala, and
  Zapperi}}]{Alava-AdP06}
\bibinfo{author}{\bibfnamefont{M.~J.} \bibnamefont{Alava}},
  \bibinfo{author}{\bibfnamefont{P.}~\bibnamefont{Nukala}}, \bibnamefont{and}
  \bibinfo{author}{\bibfnamefont{S.}~\bibnamefont{Zapperi}},
  \bibinfo{journal}{Adv. Phys.} \textbf{\bibinfo{volume}{55}},
  \bibinfo{pages}{349} (\bibinfo{year}{2006}).

\bibitem[{\citenamefont{Bouchaud et~al.}(1990)\citenamefont{Bouchaud, Lapasset,
  and Plan\`es}}]{EBouchaud90EPL}
\bibinfo{author}{\bibfnamefont{E.}~\bibnamefont{Bouchaud}},
  \bibinfo{author}{\bibfnamefont{G.}~\bibnamefont{Lapasset}}, \bibnamefont{and}
  \bibinfo{author}{\bibfnamefont{J.}~\bibnamefont{Plan\`es}},
  \bibinfo{journal}{Europhys. Lett.} \textbf{\bibinfo{volume}{13}},
  \bibinfo{pages}{73} (\bibinfo{year}{1990}).

\bibitem[{\citenamefont{M{\aa}l{\o}y et~al.}(1992)\citenamefont{M{\aa}l{\o}y,
  Hansen, Hinrichsen, and Roux}}]{Maloy-PRL92}
\bibinfo{author}{\bibfnamefont{K.~J.} \bibnamefont{M{\aa}l{\o}y}},
  \bibinfo{author}{\bibfnamefont{A.}~\bibnamefont{Hansen}},
  \bibinfo{author}{\bibfnamefont{E.~L.} \bibnamefont{Hinrichsen}},
  \bibnamefont{and} \bibinfo{author}{\bibfnamefont{S.}~\bibnamefont{Roux}},
  \bibinfo{journal}{Phys. Rev. Lett.} \textbf{\bibinfo{volume}{68}},
  \bibinfo{pages}{213} (\bibinfo{year}{1992}).

\bibitem[{\citenamefont{Schmittbuhl and
  M{\aa}l{\o}y}(1997)}]{Schmittbuhl-PRL97}
\bibinfo{author}{\bibfnamefont{J.}~\bibnamefont{Schmittbuhl}} \bibnamefont{and}
  \bibinfo{author}{\bibfnamefont{K.~J.} \bibnamefont{M{\aa}l{\o}y}},
  \bibinfo{journal}{Phys. Rev. Lett.} \textbf{\bibinfo{volume}{78}},
  \bibinfo{pages}{3888} (\bibinfo{year}{1997}).

\bibitem[{\citenamefont{Boffa et~al.}(1998)\citenamefont{Boffa, Allain, and
  Hulin}}]{Boffa-EPJAP98}
\bibinfo{author}{\bibfnamefont{J.~M.} \bibnamefont{Boffa}},
  \bibinfo{author}{\bibfnamefont{C.}~\bibnamefont{Allain}}, \bibnamefont{and}
  \bibinfo{author}{\bibfnamefont{J.-P.} \bibnamefont{Hulin}},
  \bibinfo{journal}{Eur. Phys. J. Appl. Phys.} \textbf{\bibinfo{volume}{2}},
  \bibinfo{pages}{281} (\bibinfo{year}{1998}).

\bibitem[{\citenamefont{Hinojosa and Aldaco}(2002)}]{Hinojosa-JMR02}
\bibinfo{author}{\bibfnamefont{M.}~\bibnamefont{Hinojosa}} \bibnamefont{and}
  \bibinfo{author}{\bibfnamefont{J.}~\bibnamefont{Aldaco}},
  \bibinfo{journal}{J. Mat. Res.} \textbf{\bibinfo{volume}{17}},
  \bibinfo{pages}{1276} (\bibinfo{year}{2002}).

\bibitem[{\citenamefont{Bonamy et~al.}(2006{\natexlab{a}})\citenamefont{Bonamy,
  Ponson, Prades, Bouchaud, and Guillot}}]{Bonamy-PRL06}
\bibinfo{author}{\bibfnamefont{D.}~\bibnamefont{Bonamy}},
  \bibinfo{author}{\bibfnamefont{L.}~\bibnamefont{Ponson}},
  \bibinfo{author}{\bibfnamefont{S.}~\bibnamefont{Prades}},
  \bibinfo{author}{\bibfnamefont{E.}~\bibnamefont{Bouchaud}}, \bibnamefont{and}
  \bibinfo{author}{\bibfnamefont{C.}~\bibnamefont{Guillot}},
  \bibinfo{journal}{Phys. Rev. Lett.} \textbf{\bibinfo{volume}{97}},
  \bibinfo{pages}{135504} (\bibinfo{year}{2006}{\natexlab{a}}).

\bibitem[{\citenamefont{Ponson et~al.}(2006)\citenamefont{Ponson, Auradou,
  Vi\'e, and Hulin}}]{Ponson-PRL06b}
\bibinfo{author}{\bibfnamefont{L.}~\bibnamefont{Ponson}},
  \bibinfo{author}{\bibfnamefont{H.}~\bibnamefont{Auradou}},
  \bibinfo{author}{\bibfnamefont{P.}~\bibnamefont{Vi\'e}}, \bibnamefont{and}
  \bibinfo{author}{\bibfnamefont{J.-P.} \bibnamefont{Hulin}},
  \bibinfo{journal}{Phys. Rev. Lett.} \textbf{\bibinfo{volume}{97}},
  \bibinfo{pages}{125501} (\bibinfo{year}{2006}).

\bibitem[{\citenamefont{Schmittbuhl et~al.}(1995)\citenamefont{Schmittbuhl,
  Roux, Vilotte, and M{\aa}l{\o}y}}]{Schmittbuhl-PRL95}
\bibinfo{author}{\bibfnamefont{J.}~\bibnamefont{Schmittbuhl}},
  \bibinfo{author}{\bibfnamefont{S.}~\bibnamefont{Roux}},
  \bibinfo{author}{\bibfnamefont{J.~P.} \bibnamefont{Vilotte}},
  \bibnamefont{and} \bibinfo{author}{\bibfnamefont{K.~J.}
  \bibnamefont{M{\aa}l{\o}y}}, \bibinfo{journal}{Phys. Rev. Lett.}
  \textbf{\bibinfo{volume}{74}}, \bibinfo{pages}{1787} (\bibinfo{year}{1995}).

\bibitem[{\citenamefont{Kardar}(1998)}]{Kardar-PR98}
\bibinfo{author}{\bibfnamefont{M.}~\bibnamefont{Kardar}},
  \bibinfo{journal}{Phys. Rep.} \textbf{\bibinfo{volume}{301}},
  \bibinfo{pages}{85} (\bibinfo{year}{1998}).

\bibitem[{\citenamefont{Gao and Rice}(1989)}]{GaoRice-JAM89}
\bibinfo{author}{\bibfnamefont{H.}~\bibnamefont{Gao}} \bibnamefont{and}
  \bibinfo{author}{\bibfnamefont{J.~R.} \bibnamefont{Rice}},
  \bibinfo{journal}{J. Appl. Mech.} \textbf{\bibinfo{volume}{56}},
  \bibinfo{pages}{828} (\bibinfo{year}{1989}).

\bibitem[{\citenamefont{Willis and Movchan}(1997)}]{Willis-JMPS97}
\bibinfo{author}{\bibfnamefont{J.~R.} \bibnamefont{Willis}} \bibnamefont{and}
  \bibinfo{author}{\bibfnamefont{A.~B.} \bibnamefont{Movchan}},
  \bibinfo{journal}{J. Phys. Mech. Sol.} \textbf{\bibinfo{volume}{45}},
  \bibinfo{pages}{591} (\bibinfo{year}{1997}).

\bibitem[{\citenamefont{Daguier et~al.}(1997)\citenamefont{Daguier, Nghi\^em,
  Bouchaud, and Creuzet}}]{Daguier-PRL97}
\bibinfo{author}{\bibfnamefont{P.}~\bibnamefont{Daguier}},
  \bibinfo{author}{\bibfnamefont{B.}~\bibnamefont{Nghi\^em}},
  \bibinfo{author}{\bibfnamefont{E.}~\bibnamefont{Bouchaud}}, \bibnamefont{and}
  \bibinfo{author}{\bibfnamefont{F.}~\bibnamefont{Creuzet}},
  \bibinfo{journal}{Phys. Rev. Lett.} \textbf{\bibinfo{volume}{78}},
  \bibinfo{pages}{1062} (\bibinfo{year}{1997}).

\bibitem[{\citenamefont{Rosso and Krauth}(2002)}]{Rosso-PRE02}
\bibinfo{author}{\bibfnamefont{A.}~\bibnamefont{Rosso}} \bibnamefont{and}
  \bibinfo{author}{\bibfnamefont{W.}~\bibnamefont{Krauth}},
  \bibinfo{journal}{Phys. Rev. E} \textbf{\bibinfo{volume}{65}},
  \bibinfo{pages}{025101(R)} (\bibinfo{year}{2002}).

\bibitem[{\citenamefont{Vandembroucq and Roux}(2004)}]{VR-PRE04}
\bibinfo{author}{\bibfnamefont{D.}~\bibnamefont{Vandembroucq}}
  \bibnamefont{and} \bibinfo{author}{\bibfnamefont{S.}~\bibnamefont{Roux}},
  \bibinfo{journal}{Phys. Rev. E} \textbf{\bibinfo{volume}{70}},
  \bibinfo{pages}{026103} (\bibinfo{year}{2004}).

\bibitem[{\citenamefont{Dalmas et~al.}(2007)\citenamefont{Dalmas, Lelarge, and
  Vandembroucq}}]{DLV-JNCS07}
\bibinfo{author}{\bibfnamefont{D.}~\bibnamefont{Dalmas}},
  \bibinfo{author}{\bibfnamefont{A.}~\bibnamefont{Lelarge}}, \bibnamefont{and}
  \bibinfo{author}{\bibfnamefont{D.}~\bibnamefont{Vandembroucq}},
  \bibinfo{journal}{J. Non-Cryst. Solids} \textbf{\bibinfo{volume}{353}},
  \bibinfo{pages}{4672} (\bibinfo{year}{2007}).

\bibitem[{\citenamefont{Sarlat et~al.}(2006)\citenamefont{Sarlat, Lelarge,
  S{\o}nderg{\aa}rd, and Vandembroucq}}]{SLSV-EPJB06}
\bibinfo{author}{\bibfnamefont{T.}~\bibnamefont{Sarlat}},
  \bibinfo{author}{\bibfnamefont{A.}~\bibnamefont{Lelarge}},
  \bibinfo{author}{\bibfnamefont{E.}~\bibnamefont{S{\o}nderg{\aa}rd}},
  \bibnamefont{and}
  \bibinfo{author}{\bibfnamefont{D.}~\bibnamefont{Vandembroucq}},
  \bibinfo{journal}{Eur. Phys. J. B} \textbf{\bibinfo{volume}{54}},
  \bibinfo{pages}{121} (\bibinfo{year}{2006}).

\bibitem[{\citenamefont{Ramanathan}(1998)}]{Ramanathan-PhD98}
\bibinfo{author}{\bibfnamefont{S.}~\bibnamefont{Ramanathan}}, Ph.D. thesis,
  \bibinfo{school}{Harvard University} (\bibinfo{year}{1998}).

\bibitem[{\citenamefont{Adda-Bedia et~al.}(2006)\citenamefont{Adda-Bedia,
  Katzav, and Vandembroucq}}]{AKV-PRE06}
\bibinfo{author}{\bibfnamefont{M.}~\bibnamefont{Adda-Bedia}},
  \bibinfo{author}{\bibfnamefont{E.}~\bibnamefont{Katzav}}, \bibnamefont{and}
  \bibinfo{author}{\bibfnamefont{D.}~\bibnamefont{Vandembroucq}},
  \bibinfo{journal}{Phys. Rev. E} \textbf{\bibinfo{volume}{73}},
  \bibinfo{pages}{035106(R)} (\bibinfo{year}{2006}).

\bibitem[{\citenamefont{Katzav and Adda-Bedia}(2006)}]{Katzav-EPL06}
\bibinfo{author}{\bibfnamefont{E.}~\bibnamefont{Katzav}} \bibnamefont{and}
  \bibinfo{author}{\bibfnamefont{M.}~\bibnamefont{Adda-Bedia}},
  \bibinfo{journal}{Europhys. Lett.} \textbf{\bibinfo{volume}{76}},
  \bibinfo{pages}{450} (\bibinfo{year}{2006}).

\bibitem[{\citenamefont{Bonamy et~al.}(2006{\natexlab{b}})\citenamefont{Bonamy,
  Prades, Rountree, Ponson, Dalmas, Bouchaud, Ravi-Chandar, and
  Guillot}}]{Bonamy-IJF06}
\bibinfo{author}{\bibfnamefont{D.}~\bibnamefont{Bonamy}},
  \bibinfo{author}{\bibfnamefont{S.}~\bibnamefont{Prades}},
  \bibinfo{author}{\bibfnamefont{C.~L.} \bibnamefont{Rountree}},
  \bibinfo{author}{\bibfnamefont{L.}~\bibnamefont{Ponson}},
  \bibinfo{author}{\bibfnamefont{D.}~\bibnamefont{Dalmas}},
  \bibinfo{author}{\bibfnamefont{E.}~\bibnamefont{Bouchaud}},
  \bibinfo{author}{\bibfnamefont{K.}~\bibnamefont{Ravi-Chandar}},
  \bibnamefont{and} \bibinfo{author}{\bibfnamefont{C.}~\bibnamefont{Guillot}},
  \bibinfo{journal}{Int. J. Fract.} \textbf{\bibinfo{volume}{140}},
  \bibinfo{pages}{3} (\bibinfo{year}{2006}{\natexlab{b}}).

\bibitem[{\citenamefont{Bouchbinder et~al.}(2007)\citenamefont{Bouchbinder,
  Bregman, and Procaccia}}]{Bouchbinder-PRE07}
\bibinfo{author}{\bibfnamefont{E.}~\bibnamefont{Bouchbinder}},
  \bibinfo{author}{\bibfnamefont{M.}~\bibnamefont{Bregman}}, \bibnamefont{and}
  \bibinfo{author}{\bibfnamefont{I.}~\bibnamefont{Procaccia}},
  \bibinfo{journal}{Phys. Rev. E} \textbf{\bibinfo{volume}{76}},
  \bibinfo{pages}{025101(R)} (\bibinfo{year}{2007}).

\bibitem[{\citenamefont{Hansen and Schmittbuhl}(2003)}]{Hansen-PRL03}
\bibinfo{author}{\bibfnamefont{A.}~\bibnamefont{Hansen}} \bibnamefont{and}
  \bibinfo{author}{\bibfnamefont{J.}~\bibnamefont{Schmittbuhl}},
  \bibinfo{journal}{Phys. Rev. Lett.} \textbf{\bibinfo{volume}{90}},
  \bibinfo{pages}{045504} (\bibinfo{year}{2003}).

\bibitem[{\citenamefont{Schmittbuhl et~al.}(2003)\citenamefont{Schmittbuhl,
  Hansen, and Batrouni}}]{Schmittbuhl-PRL03}
\bibinfo{author}{\bibfnamefont{J.}~\bibnamefont{Schmittbuhl}},
  \bibinfo{author}{\bibfnamefont{A.}~\bibnamefont{Hansen}}, \bibnamefont{and}
  \bibinfo{author}{\bibfnamefont{G.~G.} \bibnamefont{Batrouni}},
  \bibinfo{journal}{Phys. Rev. Lett.} \textbf{\bibinfo{volume}{90}},
  \bibinfo{pages}{045505} (\bibinfo{year}{2003}).

\bibitem[{\citenamefont{Alava and Zapperi}(2004)}]{Alava-PRL04}
\bibinfo{author}{\bibfnamefont{M.~J.} \bibnamefont{Alava}} \bibnamefont{and}
  \bibinfo{author}{\bibfnamefont{S.}~\bibnamefont{Zapperi}},
  \bibinfo{journal}{Phys. Rev. Lett.} \textbf{\bibinfo{volume}{92}},
  \bibinfo{pages}{049601} (\bibinfo{year}{2004}).

\bibitem[{\citenamefont{Schmittbuhl et~al.}(2004)\citenamefont{Schmittbuhl,
  Hansen, and Batrouni}}]{Schmittbuhl-PRL04}
\bibinfo{author}{\bibfnamefont{J.}~\bibnamefont{Schmittbuhl}},
  \bibinfo{author}{\bibfnamefont{A.}~\bibnamefont{Hansen}}, \bibnamefont{and}
  \bibinfo{author}{\bibfnamefont{G.~G.} \bibnamefont{Batrouni}},
  \bibinfo{journal}{Phys. Rev. Lett.} \textbf{\bibinfo{volume}{92}},
  \bibinfo{pages}{049602} (\bibinfo{year}{2004}).

\bibitem[{\citenamefont{C\'elari\'e et~al.}(2003)\citenamefont{C\'elari\'e,
  Prades, Bonamy, Ferrero, Bouchaud, Guillot, and Marli\`ere}}]{Celarie-PRL03}
\bibinfo{author}{\bibfnamefont{F.}~\bibnamefont{C\'elari\'e}},
  \bibinfo{author}{\bibfnamefont{S.}~\bibnamefont{Prades}},
  \bibinfo{author}{\bibfnamefont{D.}~\bibnamefont{Bonamy}},
  \bibinfo{author}{\bibfnamefont{L.}~\bibnamefont{Ferrero}},
  \bibinfo{author}{\bibfnamefont{E.}~\bibnamefont{Bouchaud}},
  \bibinfo{author}{\bibfnamefont{C.}~\bibnamefont{Guillot}}, \bibnamefont{and}
  \bibinfo{author}{\bibfnamefont{C.}~\bibnamefont{Marli\`ere}},
  \bibinfo{journal}{Phys. Rev. Lett.} \textbf{\bibinfo{volume}{90}},
  \bibinfo{pages}{075504} (\bibinfo{year}{2003}).

\bibitem[{\citenamefont{Guin and Wiederhorn}(2006)}]{Guin-IJF06}
\bibinfo{author}{\bibfnamefont{J.-P.} \bibnamefont{Guin}} \bibnamefont{and}
  \bibinfo{author}{\bibfnamefont{S.}~\bibnamefont{Wiederhorn}},
  \bibinfo{journal}{Int. J. Fract.} \textbf{\bibinfo{volume}{140}},
  \bibinfo{pages}{26} (\bibinfo{year}{2006}).

\bibitem[{\citenamefont{Roux et~al.}(2003)\citenamefont{Roux, Vandembroucq, and
  Hild}}]{RVH-EJMA03}
\bibinfo{author}{\bibfnamefont{S.}~\bibnamefont{Roux}},
  \bibinfo{author}{\bibfnamefont{D.}~\bibnamefont{Vandembroucq}},
  \bibnamefont{and} \bibinfo{author}{\bibfnamefont{F.}~\bibnamefont{Hild}},
  \bibinfo{journal}{Eur. J. Mech. A} \textbf{\bibinfo{volume}{22}},
  \bibinfo{pages}{743} (\bibinfo{year}{2003}).

\end{thebibliography}


\vfill

\newpage

\end{document}